\documentclass[manuscript,screen,nonacm]{acmart}

\AtBeginDocument{%
  \providecommand\BibTeX{{%
    \normalfont B\kern-0.5em{\scshape i\kern-0.25em b}\kern-0.8em\TeX}}}

\renewcommand\footnotetextcopyrightpermission[1]{} 
\begin{document}

\title{VedicViz: Towards Visualizing Vedic Principles in Mental Arithmetic}

\author{Noble Saji Mathews, Akhila Sri Manasa Venigalla and Sridhar Chimalakonda}

\affiliation{%
  \institution{\\Research in Intelligent Software \& Human Analytics (RISHA) Lab\\
  Department of Computer Science and Engineering\\
  Indian Institute of Technology Tirupati}
  \city{Tirupati}
  \country{India}
}
\email{{ch19b023, cs19d504, ch}@iittp.ac.in}

\begin{abstract}
  Augmenting teaching with visualization can help students understand concepts better. Researchers have leveraged visualization to teach conventional mathematics some examples being spatial and origami visualizations. 
Apart from conventional mathematics, systems such as mental arithmetic involve techniques for rapid calculation without the use of any computing tools and hence have been used in developing computational competence among students.
Vedic Mathematics is one such set of techniques for mental computation. 
However, there is a lack of technical tools which tackle mental arithmetic concepts and provide aid in the teaching of these topics to school students.
Therefore, we propose \textit{VedicViz}, a web portal that provides dynamic visualization of mathematical operations such as addition, multiplication and square root calculation, based on techniques in Vedic Mathematics. The web portal also provides visualization that enables learners to compare and contrast the mental mathematics based approach with the traditional methods for various inputs and operations. 
We evaluated \textit{VedicViz} with 20 volunteers, who were in their high school education level. They found our web portal to be useful in practicing and learning to use the methods to perform various mathematical operations.
\end{abstract}


\maketitle
\thispagestyle{empty}
\section{Introduction}

Visualization has been augmented with the teaching of several concepts in multiple domains such as physics, programming, mathematics and literature to name a few \citep{shatri2017use,teichrew2020augmented,fuchsova2019visualisation, geng2020augmented}. 
Teaching various algorithms in programming domains such as sorting algorithms through visualization has been observed to improve concept acquisition over non-visualized teaching \citep{shaziya2021strategies}. Various mathematical concepts such as geometry and linear algebra have also been taught using visualization to enhance conceptual learning and ease the level of understanding \citep{presmeg2020visualization,misrom2020enhancing,donevska2018fostering}.

Mental mathematics is an area which has been explored as way of developing ‘relational understanding’ of mathematics \citep{ineson2020mental}. It involves performing calculations in ones own head without the support of any tools. Vedic Mathematics is one such system which is popular in India \citep{raikhola2020thematic,shriki2018engagement}. 
Vedic Mathematics principles have been used in various scenarios to reduce computation times \citep{fernandes2013application}. Faster additions and multiplications are of extreme importance in Digital Signal Processing, especially in designing convolutions and digital filters \citep{huddar2013novel, arunachalam2015implementation}. 
Sushma et al. describe the use of ancient Vedic Mathematics techniques to develop a high-speed compressor based multiplier \citep{huddar2013novel}. This approach turned out to be nearly two times faster than the popular methods of multiplication during their evaluation \citep{huddar2013novel}. Arunachalam et al. describe the implementation of fast fourier transforms (FFT) using Vedic Mathematics algorithms, which had significant performance improvements as compared to conventional FFT \citep{arunachalam2015implementation}. Vedic principles have also been implemented on a FPGA to reduce delays in square and cube computations of binary numbers \citep{kodali2015fpga}. Knowledge of vedic systems ad other ancient Indian scriptures is also considered to contribute to better meta-cognitive learning capabilities \cite{chaturvedi2022differential}.

According to Mudali et. al. proper graphical visualisation is essential for students to comprehend mathematics and their ability is dependent on the media used by their teachers for visualisation \citep{mudaly2010role}. Even though Vedic Mathematics is being used for various purposes, there are no approaches that use visualization to teach Vedic Mathematics or other systems of mental mathematics in the literature like the Trachtenberg system \citep{trachtenberg2011trachtenberg}. 
Existing ways of teaching mental arithmetic are limited to reading texts 
and video tutorials. These can be supplemented with visualizations to improve the understandability and comprehension of Vedic concepts.

Vedic Mathematics finds application in not only the arithmetic domain, but also in algebra, geometry, trigonometry, calculus and applied mathematics by supplementing the usual methods and helping speed up calculations done mentally \citep{tirtha1992vedic}. Hence, we propose \textit{VedicViz}, to help teach mental arithmetic through visualizations. We take the example of Vedic Mathematics System in this paper. \textit{VedicViz} enables users to perform the desired computation for the selected mathematical operation depicting both traditional and mental mathematics workflows. In this paper we describe its use towards visualizing Vedic Mathematics as an example of the same. We used feedback from school math teachers while building the tool and evaluated it with high school students. The primary of objective of \textit{VedicViz} is to aid students learn the methods involved in mental arithmetic systems such as \textit{sutras} in Vedic Mathematics using dynamic visualizations. The usefulness of \textit{VedicViz} as a teaching aid was validated by surveying high school level students and their teachers who are the main target audience for this tool.

\section{Related Work}

Researchers have adopted various teaching aids and specifically visualizations to improve quality of teaching and learning in multiple fields \citep{fadiran2018can, asokhia2009improvisation, jacquesson2020stereoscopic, bouvska2017opportunities}. We can also observe a rise in interest in utilisation of visualisation to teach students \citep{motschnig2016team} and also Beyer et. al. have attempted teaching visualisation online \citep{beyer2016teaching}. A few examples of this include teaching engineering students abstract mathematical concepts using simulations and animations \citep{hadjerrouit2018using}, in medicine for explaining the anatomy of the brain and the skull through stereoscopic 3D visualization \citep{jacquesson2020stereoscopic} and even towards enhancing coordination of construction projects \citep{liapi20034d, bouvska2017opportunities}. Visualization-based technologies have also found widespread use in the teaching domain, as mentioned above, since using them as Teaching aids to assist the learning experience has helped learners understand various concepts easily \citep{ye2007enhancing,firat2018towards}.

Visualization is central to learning subjects where students are expected to be meta-cognitive with respect to visualizations such as the core sciences \citep{parish2019data}. External representations (diagrams, graphs, and representative models) are considered to be valuable tools for teachers \citep{bouvska2017opportunities,jacquesson2020stereoscopic}. The formation of various visualizations using these representations plays a key role in multiple areas of learning and helps students understand and recollect concepts better \citep{gilbert2005visualization}.

Digital intervention in teaching and learning of mathematics is an area that has garnered interest from researchers and practitioners alike \citep{borba2017digital}. 
The use of interactive technology to create mathematical models has been a subject of study to help build upon objects that are used towards aiding with abstractions \citep{bishop1989review}. Other studies have also compared flipper learning models with gamification to traditional methods \citep{lo2020comparison}. The findings show that the approach promoted cognitive engagement among students. Considering the impact of gamification, Kamalodeen et al.,  have presented the methods involved in including gamification elements to teach geometry in elementary schools \citep{kamalodeen2021designing}. Keegan et al. describe use of virtual reality based visualizations to aid in teaching multi-variable calculus \citep{kang2020impact}, however this did not translate to better understanding. The authors find that VR is not  always a suitable replacement to classroom based learning but can be used as teaching aids. Construct3D is a tool designed for mathematics and geometry education \citep{kaufmann2002construct3d}. It helps with 3D geometry construction and is based on "Studierstube", which is a collaborative augmented reality system. The study describes efforts towards how the tool helps in maximizing the transfer of learning. \textit{TrueBiters} has been proposed as an educational game to help students in the first undergraduate course learn and practice concepts in propositional logic \citep{de2019truebiters}. This game is designed as a two-player mobile application and motivates students in practicing truth-tables corresponding to different logical operators in propositional logic \citep{de2019truebiters}.
 
Few studies also explore the use of open source tools such as GeoGebra \footnote{https://www.geogebra.org/}, an interactive application that aims to help  visualize concepts in mathematics and science and Desmos\footnote{https://www.desmos.com/}, a tool which can quickly create graphical simulations for equations. One of the works in literature also describes use of the tool to support high school students’ understanding of conditional probability and Bayes’ theorem \citep{aizikovitsh2012teaching}. The work also aims to improve high school students’ risk literacy and critical thinking. SimReal is a visualization tool for teaching mathematics \citep{hadjerrouit2018using}. It provides video lessons, interactive simulations, and animations to enhance comprehension and engagement in abstract mathematics subjects. Güven et al. highlight the potential of a dynamic geometry software (DGS), Cabri 3D to teach analytic geometry of space \citep{guven2008effect}. The paper also states how prospective mathematics teachers found the tool helpful in simplifying the explanation of 3D geometry concepts to students.

 
Visualization has gained increased visibility in the domain of mathematics education. It plays a rich role in helping out students in learning and doing mathematics \citep{presmeg2020visualization}. Roderick Nicolson describes \textit{SUMIT} a system that assists in teaching of arithmetic in the classroom \citep{nicolson1990design}.  
The positive impacts of visualization based learning in mathematics as discussed above and the lack of tools that better aid in learning mental mathematics systems serve as the motivation for this work.  A majority of the tech based interventions in mathematics education have been classified as augmentation or that it aims to improve the traditional approaches through functional development \citep{bray2017technology}. Along similar lines we propose \textit{VedicViz} as a tool to aid learners in understanding and practicing many useful techniques of Vedic Mathematics, through the use of visualizations.

\section{Design and Development of \textit{VedicViz}}


The current version of \textit{VedicViz} supports simulating addition, multiplication and square root evaluation based on some of the 16 main Sutras/formulae in Vedic Mathematics. The Vedic Mathematics system also includes 13 more sub-sutras and their corollaries, along with the main sutras. One of these formulae is used as a part of \textit{VedicViz}, the \textit{URDHVA-TRIBHAGYAM}, which is Sanskrit for \textit{vertically and crosswise}. 
This method makes use of cross multiplications to directly get digits in the final answer without necessitating a full evaluation. Other techniques used in the visualizer include place value-based addition, which is provided in the first level as it is easy to understand and implement. For square roots, an analytic approach is implemented. One of the issues with a large number of rapid math tricks and Vedic techniques is that there are usually restrictions on the cases where they can be used. The methods had to be selected to address this issue so that they are not too context-specific. This information is further incorporated into the information about each method that is being demonstrated by \textit{VedicViz}. In the case of operands for which the method cannot be run, a warning message will be displayed.

\begin{figure}
\includegraphics[width=\linewidth]{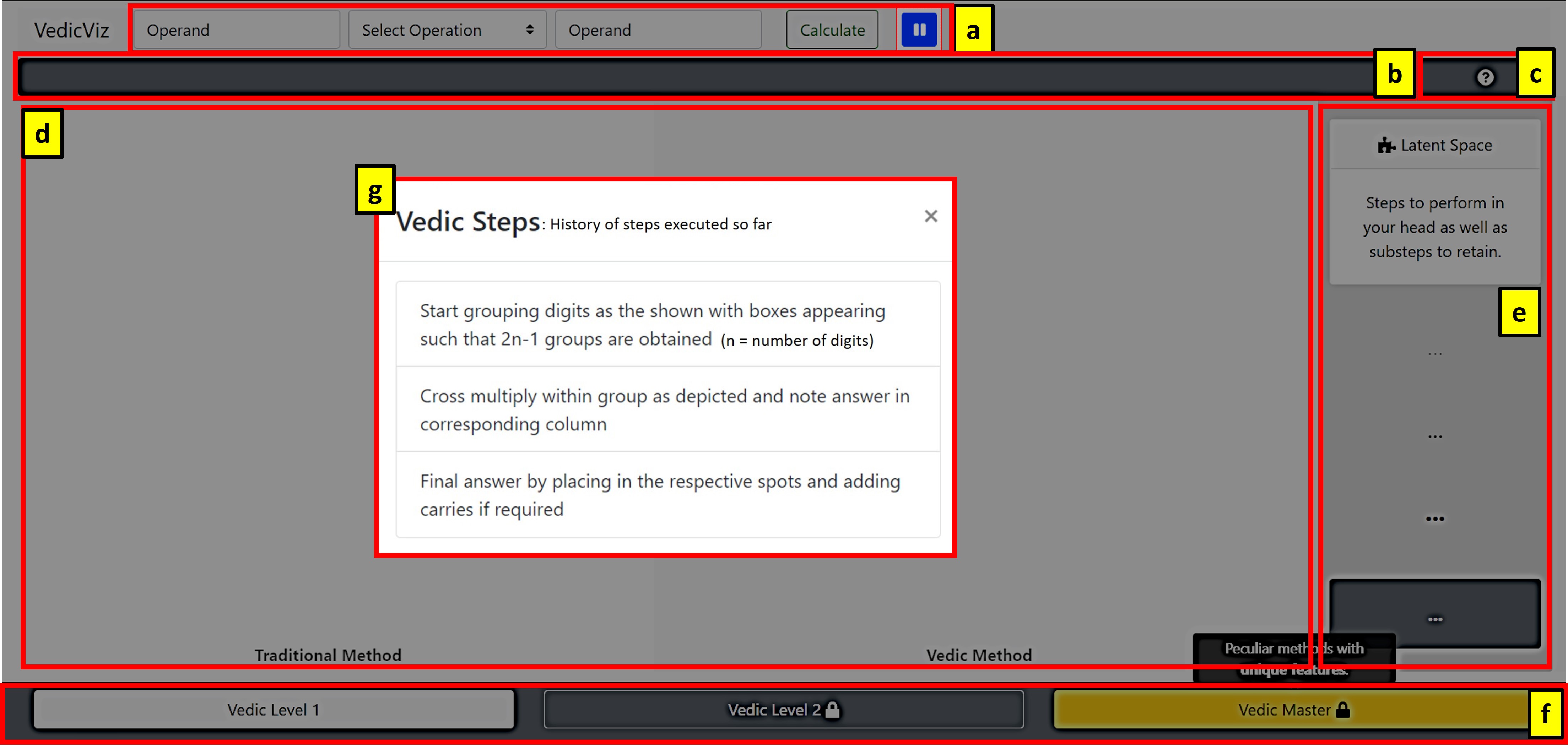}
\caption{Front-end components and controls: (a) Control components to begin and pause visualization, (b) Live rolling list of steps followed, (c) Information about selected method, (d) Dual visualization area, (e) Latent Space for sub-steps, (f) Method options, (g) History of executed steps revealed on tapping component c.} 
\label{fig:breakup}
\end{figure}

\subsection{Web App Structure}

\textit{VedicViz} is a web deployed React application \footnote{https://reactjs.org/} that makes use of the flexibility available in the framework to allow visualization of Vedic Mathematics principles in a step-wise approach. It allows the user to use its functionally as a calculator, and also it produces the computational steps involved in the calculation in a form that is replicable by the user. Components of the visualizer were built to be reusable and modular so that it can be expanded further, to support a large variety of Vedic methods. It also facilitates improving the appearance, behavior, and functionality as required. The project was initially structured into four main components, namely \textit{Control Options, Visualization Area, Latent Space} and \textit{Method switcher} as shown in Figure~\ref{fig:breakup}. The functionality of each of these was further modified based on suggestions, feedback received from high school maths teachers on improvements that would help aide teaching better. 

In the background, the flow between algorithms and the Visualizer component is handled through Async generators \footnote{https://javascript.info/async-iterators-generators}. The front-end states are managed through respective \textit{state variables}, and the publicly required information is handled through a Redux store \footnote{https://redux.js.org/api/store}. This enables invoking exposed points of entry from anywhere in the application to achieve the desired outcome. It also makes use of the react-grid-layout package \footnote{https://github.com/STRML/react-grid-layout} to support dynamic updates and resizing. The kind of layout used can be specified by the algorithm to ensure the inclusion of special blocks to be used during the visualization. This can also be utilized to have guide blocks such as the operation appender depicted in Figure~\ref{fig:breakup2}(c).

\subsection{Visualizer}

\begin{figure}
\includegraphics[width=\linewidth]{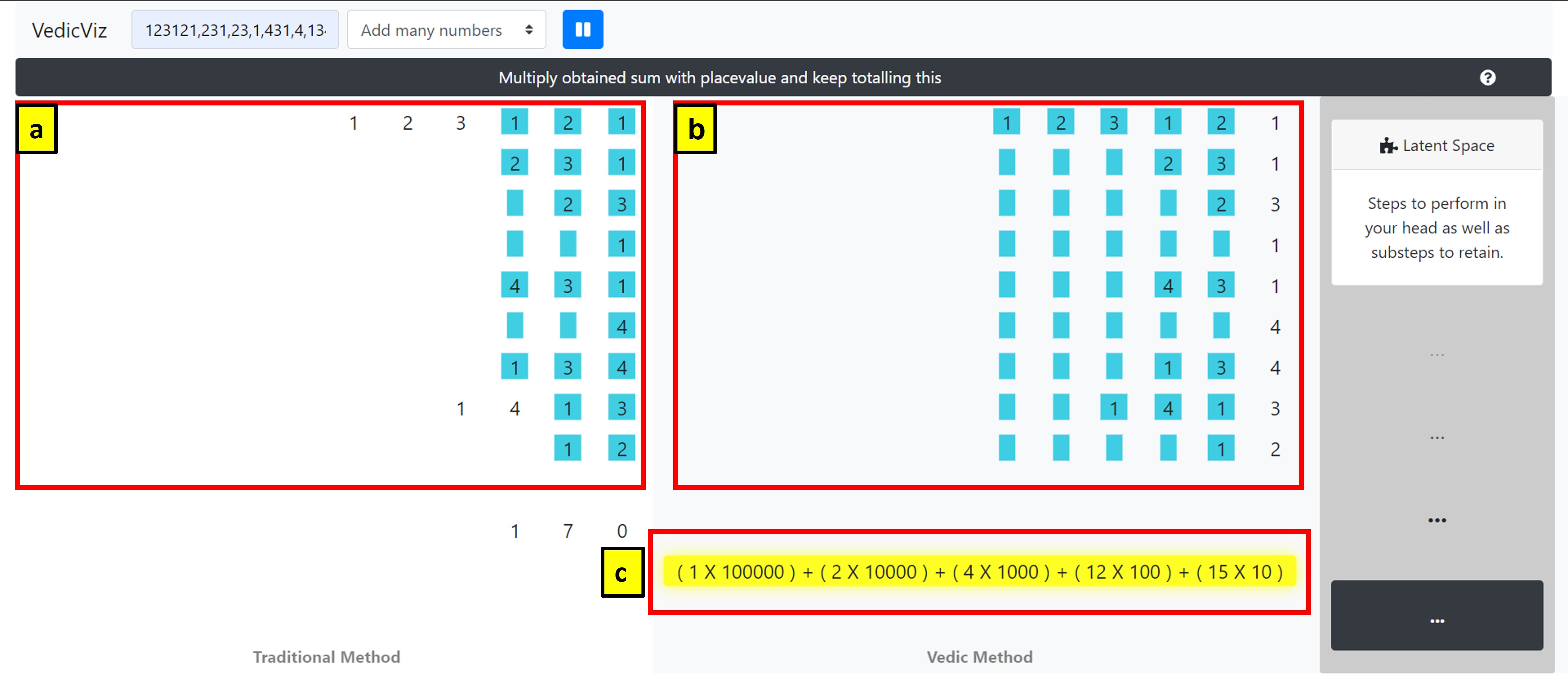}
\caption{A visualization in progress,(a) Traditional method, (b) Vedic method, (c) Custom block in action.} 
\label{fig:breakup2}
\end{figure}

The Visualizer is the primary focus of the app from an end-users' perspective. It takes input supplied after processing the respective algorithms invoked. It is divided into two areas, and the first shows the step-wise implementation of the traditional method used to solve problems, while the second area proposes an alternative Vedic method. Both components perform the same computations, however, going about it in different ways as directed by the algorithms. All implemented Vedic algorithms 
display evaluations 
similar to that meant to be performed by the students. In Figure~\ref{fig:breakup2} we see multiple numbers being added and in Figure~\ref{fig:breakup3} we show the multiplication of 2 large numbers which we will go over in detail.

In order to show the interaction between multiple cards for a step, nodes visited in each step are animated. The visited values are worked upon by the background logic implemented in the algorithm. Description of action of each step can be found in the rolling list on top. Tapping on this list displays all the steps completed so far. This can be used for review later on as well, as shown in Figure~\ref{fig:breakup}(b). 

\subsection{Latent Space}

\begin{figure}
\includegraphics[width=\linewidth]{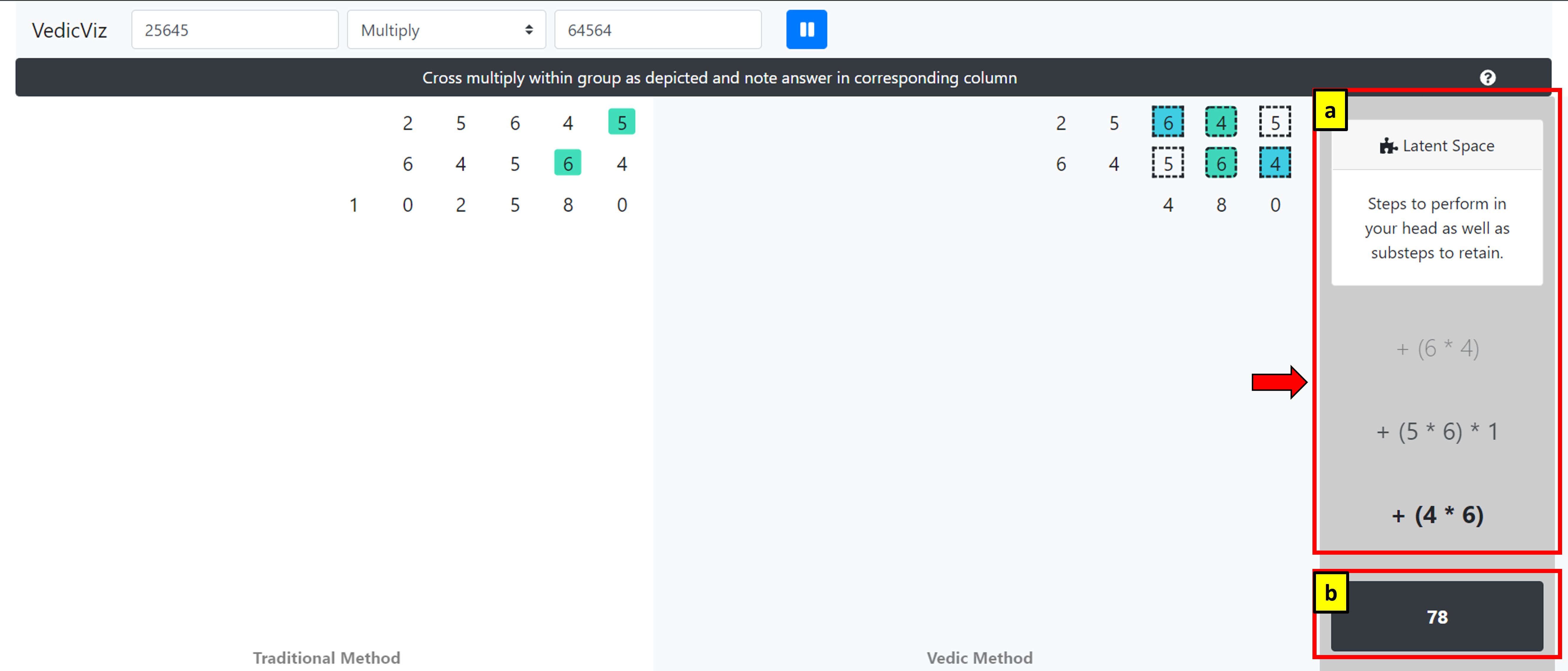}
\caption{Latent space with basic calculations being performed in (a) and evaluated in (b) for the main step.} 
\label{fig:breakup3}
\end{figure}
Each of the methods has a separate algorithm file that handles analyzing, modulating, and returning the steps to be performed onto the visualizer. Latent space, as shown in Figure~\ref{fig:breakup3}(a), displays the responses received from the Vedic algorithms. This is used to indicate the necessary basic calculations that were executed by the algorithm to get to the next step. It serves as a volatile display of the current and past two basic operations, as depicted in Figure~\ref{fig:breakup3} which shows two 5 digit numbers being multiplied.

By default, every background operation is meant to be passed through the latent space as it processes the statements sent, with the help of node package Math.js \footnote{https://mathjs.org/}, allowing for complex sub-steps if the need arises. At the end of each main step, the evaluated value is returned to the visualizer and is handled based on the method's logic. Even though both types of methods are built to pass through latent space in the current design, only the steps involved in the Vedic methods are passed through it due to the comparative layout, which could be easily enabled.

\section{User Scenario}

Consider \textit{Veda}, a high school student who wants to learn \textit{Vedic Mathematics}. On loading up the website, she is presented with the web application, its components are shown in Figure~\ref{fig:breakup}. She selects the method that she wants to learn about using the drop-down and then selects the suggested level from those available. She decides to try out the multiplication method first. On reading about the method from the info-button, indicated by  Figure~\ref{fig:breakup}(c), she learns that the method allows multiplication between numbers with an equal number of digits but is unique because the parts of the final answer are obtained at each step and hence, the complete answer is obtained in one go. The info-panel also gives her an insight into how the method can easily be extended for numbers without an equal number of digits by appending zeroes to the left. Now she enters the operands of her doubt into the app through the component depicted in Figure~\ref{fig:breakup}(a) and hits the calculate button. The visualization is displayed on the screen with a part of the main area showing the methods she was familiar with as shown by Figure~\ref{fig:breakup2}(a) and the other part depicting the Vedic principles step by step in Figure~\ref{fig:breakup2}(b).


\begin{figure}
\centering
\includegraphics[width=0.8\linewidth]{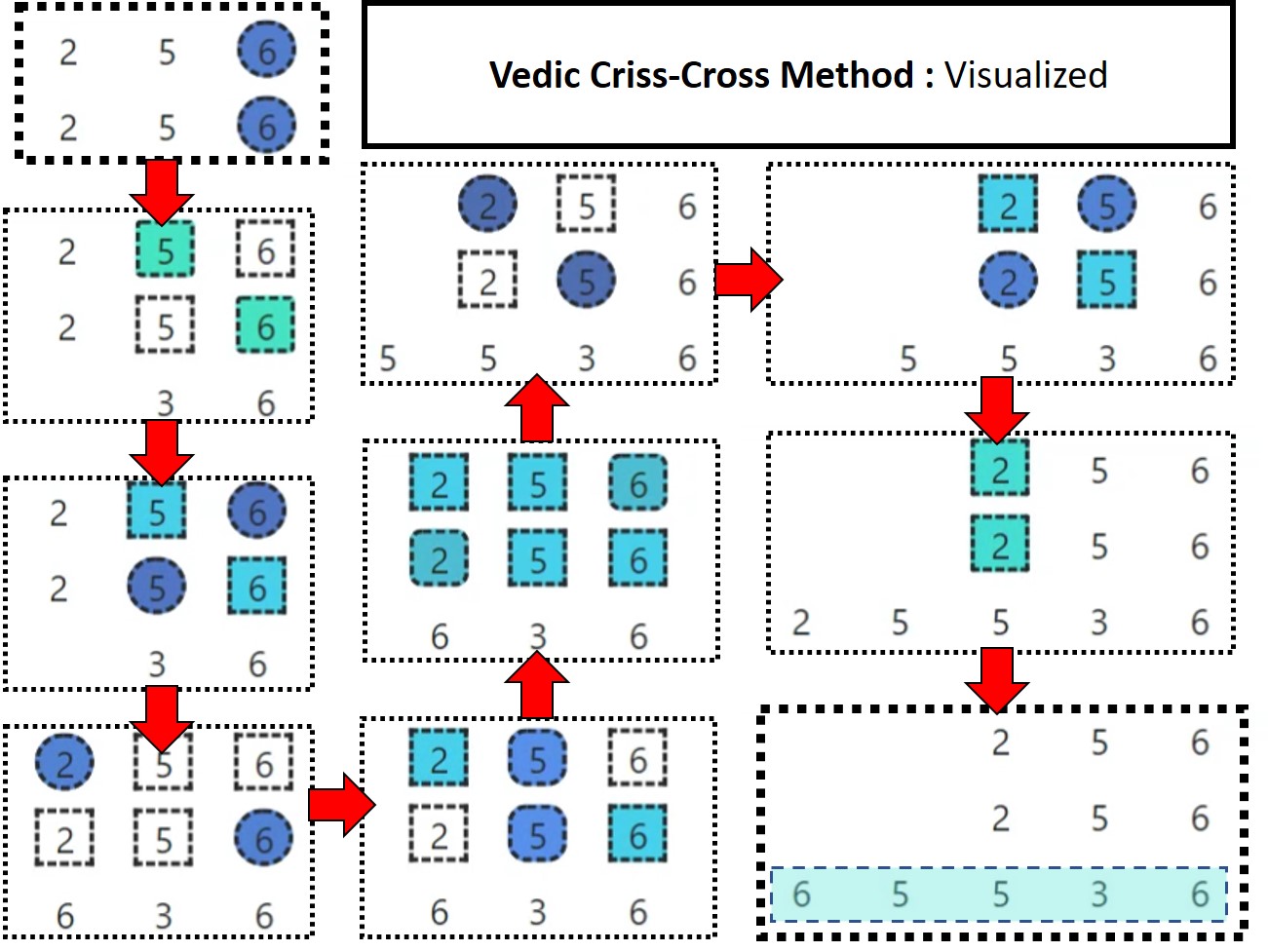}
\caption{Step-wise snapshot of Vedic criss-cross method as visualized by \textit{VedicViz}.} 
\label{fig:crisscross}
\end{figure}

The visualization of this method is shown in Figure~\ref{fig:crisscross} as a sequence of screenshots during the visualization.
The steps shown are as follows
\begin{itemize}
    \item Multiply the one’s digits together and note the answer
    \item Multiply the diagonals and add them together and note the answer considering the next place value
    \item Depending on the length of the numbers taken it will have to be repeated again.
    
    Here since its a 3 digit number we first make a group of 2 columns on the right followed by all 3 columns and finally the 2 columns on the left
    
    \item Multiply the digits in the highest place value and note the answer into the next place value
\end{itemize}
The animations are used to draw attention to the current operation being performed and changes happening on screen. The other components on the page are used in tandem to describe each of these steps as they are visualized.

 The latent memory allowed keeping track of what operations were being performed by the Vedic method. Depiction of both methods allow her to easily grasp how the methods differ from each other. She finds the main step being performed described at the top in a list, that can be opened up to view all steps executed so far. Other detailed information about the sub-steps, if required, kept appearing in a flashing yellow box. Towards the right in the latent space section, she finds all the ongoing background basic calculations as a rolling list. Once the simulation is underway, she can pause and resume, enabling her to follow along at her own pace. She also finds the ability to get steps for any specific question useful to practice and learn.

\section{Evaluation}
\textit{VedicViz} has been developed with the main aim of aiding students towards learning to use various principles known as \textit{Sutras} in \textit{Vedic Mathematics}, through visualization. Hence, we aimed to evaluate the usefulness and user experience towards using \textit{VedicViz}. Other visualization tools in the literature, developed with similar aims of helping students in learning multiple concepts across different domains, have been evaluated against similar dimensions such as user experiences, intention to use, usefulness and ease of use. Most of these evaluations included questionnaire-based user surveys, following various Technology Acceptance Models \citep{venkatesh2008technology} such as TAM (Technology Acceptance Model), UTAUT (Unified Theory of Acceptance and Use of Technology) and IDT (Innovation Diffusion Theory). Similar to the evaluation of visualization tools in the literature, \textit{VedicViz} has also been evaluated through a questionnaire-based user survey \citep{kolling2010greenfoot, yang2018evaluations}. 

Considering the factors against which \textit{VedicViz} has to be evaluated, an adapted evaluation model of TAM2 fits well, which is an enhanced version of TAM. It deals with aspects that include \textit{Perceived Ease of Use (PEOU), Perceived Usefulness (PU), Behavioral Intention to use (BI) and Actual Use (AU)}.
The decisions taken by users with respect to adapting to a new technology or innovations influence the factors mentioned in TAM2. Along with these factors, user decisions are also influenced by another factor - Complexity (C) of the technology, which we considered for evaluation of \textit{VedicViz}. 
In addition, we evaluated \textit{VedicViz} for correctness of the visualization, which refers to the accuracy and relevance of rendered visualization based on the operations selected by users. This consolidates to an enhanced and adapted TAM2 model, consisting of the following six factors 
- Perceived Ease Of Use (PEOU), Perceived Usefulness (PU), Behavioral Intention to Use (BI), Actual Use (AU), Complexity (C) and Correctness (CR).


To evaluate \textit{VedicViz}, we circulated a user survey among 20 students, who were in their high school education level. This survey consisted of a questionnaire-based on the adapted TAM2 model. The questionnaire also included questions aimed towards gathering demographic information of the participants. 
The questionnaire thus consisted of 18 questions to be answered on a five-point Likert scale, 3 of which referred to the demographics and the rest of the questions referred to user perceptions, streamlined with factors of the adapted TAM2 model. Table~\ref{tab:results} shows the questionnaire and the results obtained in terms of mean and standard deviation. 
All the volunteers were given a tutorial that included an explanation of working of \textit{VedicViz}, accompanied by a demonstration of an example visualization for sample operands and operation. After the tutorial, all volunteers were provided with the link to access \textit{VedicViz}. They were requested to use multiple options provided on \textit{VedicViz}, for various examples of their choice. The questionnaire designed by adapting the TAM2 model for \textit{VedicViz} was then forwarded to the volunteers and they were requested to answer the questionnaire-based on their experience and perception, on 5-point Likert Scale along with an option to provide open suggestions. 

We further tried to understand the usefulness of \textit{VedicViz}, as a self-learning tool, as perceived by teachers. Hence, we requested 15 school teachers to give their feedback on the tool. The questionnaire shared included the following questions:
\begin{itemize}
    \item \textit{Will the screen partitions be helpful to the students in your opinion?}
    \item \textit{Which grades do you feel this tool would be most effective in ?}
    \item \textit{How would you rate the tool for self learning ? (Rate on a scale of 1 to 5)}
\end{itemize}
Of the 13 teachers who provided feedback, 9 teach students of upper primary and 4 teach high school students. They provided their feedback in terms of what features could be added or improved in \textit{VedicViz}. 
\footnote{Survey Details \url{https://dataosf.page.link/vedicviz}}. 

\begin{table}
  \caption{Adapted TAM2 based questionnaire and its results.}
  
\begin{tabular}{|l|l|l|l|}
\hline
Factors & Questions                                                                                                                                                                                                      & Mean & SD           \\ \hline
C      & \begin{tabular}[c]{@{}l@{}}Using VedicViz is compatible with my learning style\\ (1= strongly disagree, 5 = strongly agree)\end{tabular}                                                                       & 3.6  & 0.734 \\ \hline
C      & \begin{tabular}[c]{@{}l@{}}Using VedicViz fits well with the way I like to learn \\new concepts\\ (1= strongly disagree, 5= strongly agree)\end{tabular}                                                         & 4    & 0.707 \\ \hline
PU     & \begin{tabular}[c]{@{}l@{}}Using VedicViz would improve my scope of understanding\\  real-time use of concepts in the VedicViz\\ (1= strongly disagree, 5 = strongly agree)\end{tabular}                       & 3.9  & 0.7          \\ \hline
PU     & \begin{tabular}[c]{@{}l@{}}Using VedicViz would increase my productivity in \\ learning Vedic Mathematics\\ (1= strongly disagree, 5 = strongly agree)\end{tabular}                                            & 3.95 & 0.973 \\ \hline
PU     & \begin{tabular}[c]{@{}l@{}}Using VedicViz would enhance my effectiveness in \\ learning and understanding multiple principles of \\ Vedic Mathematics\\ (1=strongly disagree, 5 = strongly agree)\end{tabular} & 4.4  & 0.489 \\ \hline
PU     & \begin{tabular}[c]{@{}l@{}}Using VedicViz would make it easier for me to \\ learn Vedic Mathematics\\ (1= strongly disagree, 5 = strongly agree)\end{tabular}                                                  & 3.9  & 0.943\\ \hline
PU    & \begin{tabular}[c]{@{}l@{}}I think using VedicViz is very useful for me to learn.\\ (1= strongly disagree, 5 = strongly agree)\end{tabular}                                                                    & 4    & 0.836 \\ \hline
PEOU   & \begin{tabular}[c]{@{}l@{}}I think learning to use VedicViz is easy\\ (1= strongly disagree, 5 = strongly agree)\end{tabular}                                                                                  & 3.75 & 0.766 \\ \hline
PEOU   & \begin{tabular}[c]{@{}l@{}}I think becoming skillful at using VedicViz is easy\\ (1= strongly disagree, 5 = strongly agree)\end{tabular}                                                                       & 4    & 0.707 \\ \hline
PEOU   & \begin{tabular}[c]{@{}l@{}}I think using VedicViz is easy\\ (1= strongly disagree, 5 = strongly agree)\end{tabular}                                                                                            & 3.95 & 0.864\\ \hline
BI     & \begin{tabular}[c]{@{}l@{}}Assuming I had access to VedicViz, I intend to use it\\ (1= strongly disagree, 5 = strongly agree)\end{tabular}                                                                     & 4.2  & 0.6          \\ \hline
BI     & \begin{tabular}[c]{@{}l@{}}VedicViz has made my learning interactive\\ (1= strongly disagree, 5 = strongly agree)\end{tabular}                                                                                 & 4.4  & 0.734 \\ \hline
AU     & \begin{tabular}[c]{@{}l@{}}I would often engage learning Vedic Mathematics \\ via VedicViz\\ (1= strongly disagree, 5 = strongly agree)\end{tabular}                                                           & 4    & 0.547 \\ \hline
CR     & \begin{tabular}[c]{@{}l@{}}Visualizations displayed by VedicViz are relevant \\ to the operations selected\\ (1= strongly disagree, 5 = strongly agree)\end{tabular}                                           & 4.2  & 0.748 \\ \hline
\end{tabular}
\label{tab:results}
\end{table}
\section{Results}
The three demographic questions in the questionnaire refers to details about the age group, gender, and familiarity with \textit{Vedic Mathematics} among the participants. The demographics-based questionnaire results indicate that the participant set consisted 35\% Female participants and 65\% Male participants, with all the participants in the age group of 16-17 years. The demographic analysis results also indicate that a majority of the participants are not familiar with \textit{Vedic Mathematics} (70\% of the participants have never used \textit{Vedic Mathematics}). Only 15\% of the participants responded to be familiar with \textit{Vedic Mathematics}. These results also indicate the need for \textit{VedicViz}, towards introducing \textit{Vedic Mathematics} to a wide range of people. The participants' common age group is due to the selected sample, as \textit{VedicViz} was aimed to help students in their higher education level in learning \textit{Vedic Mathematics}.

\begin{figure}
    \centering
    \includegraphics[width = 0.8\linewidth]{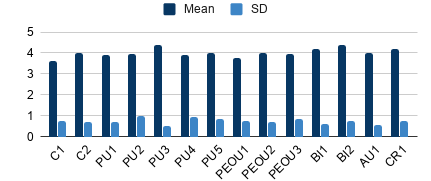}
    \caption{Mean and Standard Deviation plots for factors of adapted TAM2 model in the questionnaire.}
    \label{fig:msd}
\end{figure}

The results of the user survey for questions referring to perception and correctness, based on the adapted TAM2 model, are presented as the mean and standard deviation in Figure~\ref{fig:msd}. All the questions were answered based on a 5-point Likert Scale, where the maximum number of points(5) refers to strongly agree, while the minimum number of points(1) refers to strongly disagree. This implies that score close to 5 indicates better acceptance among the users while that close to 1 indicates lower acceptance.

The Table~\ref{tab:CorrResults} indicates correlation between the existing factors of TAM2 additional Complexity (C) factor. The factor C, is observed to have positive correlations with all the factors, PU, PEOU, BI, and AU of TAM2. The positive correlation of C with factors of TAM2 also reinforces the idea of integrating Complexity with TAM2 model. PU is observed to have a strong positive correlation with PEOU, BI, and AU, inferring that perceived usefulness is affected by ease of use, intention to use and actual use. However, a weak positive correlation of PEOU with AU indicates that perceived ease of use is scarcely affected by actual use with respect to \textit{VedicViz}. There is a strong positive correlation between BI and AU, indicating that actual use and intention are strongly dependent on each other. Table~\ref{tab:CorrResults} does not display Correctness (CR) factor as it deals with validity of \textit{VedicViz}, rather than with user decisions.

The results presented in Figure~\ref{fig:msd} indicate that all the questions have mean value greater than 3, indicating better acceptance of \textit{VedicViz} among a majority of the participants. The lower standard deviations for a majority of the questions indicate that many participants have similar views towards \textit{VedicViz}. For example, question PU3 refers to Perceived Usefulness and precisely queries the user whether ``\textit{using VedicViz would enhance users' effectiveness in learning and understanding multiple principles of Vedic Mathematics}' and has a mean value of 4.4 and standard deviation of 0.48. The maximum mean and lesser standard deviation indicates that the majority of the participants either strongly agreed or agreed to the question PU3, inferring that \textit{VedicViz} is useful to the users.

All other mean values further infer that \textit{VedicViz} is considerably easy to use, with the majority of the participants finding it useful. The mean values of questions BI1 and BI2 are greater than 4, indicating that a majority of the participants intend to use \textit{VedicViz}. Participants have also provided textual feedback and suggestions on improving \textit{VedicViz}. User feedback and suggestions include:
\begin{itemize}
    \item ``\textit{give solved examples on the website so that we can see...}"
    \item ``\textit{some references could be added so that the interested user can learn more about the topic}"
\end{itemize}

In the user survey with school teachers, 8 of the 13 teachers rated either 4 or 5 (out of 5), for self-learning through \textit{VedicViz}. They have also suggested various improvements, that included use of better animations, tutorial videos, better color-combinations, quizzes, and so on. 

\begin{table}
    \caption{Correlation Analysis Results of adapted TAM2 model.}
    {\begin{tabular}{cccccc} \toprule
     & C & PU & PEOU & BI & AU \\
     \midrule
     C & 1 & & & &\\
     PU & 0.327 & 1 & & &\\
     PEOU & 0.308 & 0.557 & 1 & &\\
     BI & 0.367 & 0.564 & 0.566 & 1 &\\
     AU & 0.484 & 0.456 & 0.152 & 0.409 & 1\\ \bottomrule
    \end{tabular}}
    \label{tab:CorrResults}
\end{table}

\section{Discussion and Limitations}

With the teaching environment continuously shifting away from the in-person instruction model, visualization techniques are becoming one of the most popular and effective means of content delivery and abstraction to better understand and enhance student learning. \textit{VedicViz} is an attempt towards the visualization of principles in \textit{Vedic Mathematics} and promoting their use in a manner that makes it easy to learn and practice.

There are certain constraints that come with the current model used for visualization. Due to the block-based approach, there are some limitations on what kinds of visualization can be implemented. Due to the large variety of Vedic methods that exist, it becomes difficult to generalize the system for all. However, since most of the methods involve manipulating the original inputs, the current workflow allows for many different ways to depict a method. Depending on the complexity of the algorithm, the amount of data to be tracked by the user might increase. Also, measures have been taken to prevent this information overload through separated out components, and specific meanings were added to updates that occur in each of them. Another limitation is the sample size of participants in the survey study. This is too small, making statistical analysis based conclusions weak.

The current version of the tool is under active development and is being improved based on feedback from learners and instructors alike. It currently incorporates four types of operations and their respective traditional implementations. However, it currently does not explain the processes in detail to satiate different target audiences. Lack of audio guidance could also make it difficult for users who are learning \textit{Vedic Mathematics} for the first time. Since the model runs on a step by step basis, it does not currently have the means to go in-depth about a particular step for long, as it would impact the overall experience on successive runs. In addition, the user interface of the tool itself could be improved to make it easy for learners.

\section{Conclusion and Future Work}
\label{conclusion}
In this paper, we have discussed the growing use of visualization techniques in the learning ecosystem employed, especially in teaching mathematics. Several visual methods have been employed in the educational domain to enhance the delivery of information and knowledge. However, to the best of our knowledge, not much work has been done towards aiding teaching of mental arithmetic systems. Hence, in this paper we have proposed a web-based application \textit{VedicViz}. The app facilitates visualization of \textit{Vedic Math} principles and compares them to the traditional workflow in order to help users in gaining a better understanding of how to use them.


The motivation behind \textit{VedicViz} was to provide a means to simplify learning and practicing mental arithmetic systems like \textit{Vedic Mathematics}. This was accomplished by building a teaching aid that allows multiple ways of visualizing math and explaining the procedure. It seeks to help promote simulation of the background mental steps in an operation, shifting the focus from direct evaluation based design to abstraction and understanding. 

\label{Future Work}
The application currently uses a dynamic table based foundation to hold components for the simulation. But this workflow introduces certain limitations to how the algorithm can be depicted. It can be extended further by building on a free canvas as a foundation to enable a large variety of visual presentations, which could also increase the models' flexibility. We are also continuing work based on received feedback in order to target different levels of expertise. Much more detailed explanations and an audio guidance system is also a part of the future work.

In the future, we plan to add specific features to make the learning experience more personalized. 
We also plan to add options to time each run and hence help the user get better with practice and by generating practice problems based on the initial query of the user. Other future plans include adding support for more methods, both numerical and analytical, into the visualizer. This includes adding new operations as well as support for a wider variety of visualizations and simplifying the existing package to facilitate easier addition of custom modules to the system to improve its functionality. 



\subsection*{Acknowledgement(s)}
We thank all the participants for their valuable time and honest feedback that helped us in evaluating \textit{VedicViz}.

\bibliographystyle{ACM-Reference-Format}
\bibliography{sample-base}


\begin{thebibliography}{45}


\ifx \showCODEN    \undefined \def \showCODEN     #1{\unskip}     \fi
\ifx \showDOI      \undefined \def \showDOI       #1{#1}\fi
\ifx \showISBNx    \undefined \def \showISBNx     #1{\unskip}     \fi
\ifx \showISBNxiii \undefined \def \showISBNxiii  #1{\unskip}     \fi
\ifx \showISSN     \undefined \def \showISSN      #1{\unskip}     \fi
\ifx \showLCCN     \undefined \def \showLCCN      #1{\unskip}     \fi
\ifx \shownote     \undefined \def \shownote      #1{#1}          \fi
\ifx \showarticletitle \undefined \def \showarticletitle #1{#1}   \fi
\ifx \showURL      \undefined \def \showURL       {\relax}        \fi
\providecommand\bibfield[2]{#2}
\providecommand\bibinfo[2]{#2}
\providecommand\natexlab[1]{#1}
\providecommand\showeprint[2][]{arXiv:#2}

\bibitem[Aizikovitsh-Udi and Radakovic(2012)]%
        {aizikovitsh2012teaching}
\bibfield{author}{\bibinfo{person}{Einav Aizikovitsh-Udi} {and}
  \bibinfo{person}{Nenad Radakovic}.} \bibinfo{year}{2012}\natexlab{}.
\newblock \showarticletitle{Teaching probability by using geogebra dynamic tool
  and implemanting critical thinking skills}.
\newblock \bibinfo{journal}{\emph{Procedia-Social and Behavioral Sciences}}
  \bibinfo{volume}{46} (\bibinfo{year}{2012}), \bibinfo{pages}{4943--4947}.
\newblock


\bibitem[Arunachalam et~al\mbox{.}(2015)]%
        {arunachalam2015implementation}
\bibfield{author}{\bibinfo{person}{S Arunachalam}, \bibinfo{person}{SM
  Khairnar}, {and} \bibinfo{person}{BS Desale}.}
  \bibinfo{year}{2015}\natexlab{}.
\newblock \showarticletitle{Implementation of Fast Fourier Transform and Vedic
  Algorithm for Image Enhancement Using Matlab}.
\newblock \bibinfo{journal}{\emph{Applied Mathematical Sciences}}
  \bibinfo{volume}{9}, \bibinfo{number}{45} (\bibinfo{year}{2015}),
  \bibinfo{pages}{2221--2234}.
\newblock


\bibitem[Asokhia(2009)]%
        {asokhia2009improvisation}
\bibfield{author}{\bibinfo{person}{MO Asokhia}.}
  \bibinfo{year}{2009}\natexlab{}.
\newblock \showarticletitle{Improvisation/teaching aids: Aid to effective
  teaching of English language}.
\newblock \bibinfo{journal}{\emph{International Journal of Educational
  Sciences}} \bibinfo{volume}{1}, \bibinfo{number}{2} (\bibinfo{year}{2009}),
  \bibinfo{pages}{79--85}.
\newblock


\bibitem[Beyer et~al\mbox{.}(2016)]%
        {beyer2016teaching}
\bibfield{author}{\bibinfo{person}{Johanna Beyer}, \bibinfo{person}{Hendrik
  Strobelt}, \bibinfo{person}{Michael Oppermann}, \bibinfo{person}{Louis
  Deslauriers}, {and} \bibinfo{person}{Hanspeter Pfister}.}
  \bibinfo{year}{2016}\natexlab{}.
\newblock \showarticletitle{Teaching visualization for large and diverse
  classes on campus and online}. In \bibinfo{booktitle}{\emph{Proceedings of
  IEEE VIS workshop on pedagogy data visualization}}.
\newblock


\bibitem[Bishop(1989)]%
        {bishop1989review}
\bibfield{author}{\bibinfo{person}{Alan~J Bishop}.}
  \bibinfo{year}{1989}\natexlab{}.
\newblock \showarticletitle{Review of research on visualization in mathematics
  education}.
\newblock \bibinfo{journal}{\emph{Focus on learning problems in mathematics}}
  \bibinfo{volume}{11}, \bibinfo{number}{1} (\bibinfo{year}{1989}),
  \bibinfo{pages}{7--16}.
\newblock


\bibitem[Borba et~al\mbox{.}(2017)]%
        {borba2017digital}
\bibfield{author}{\bibinfo{person}{Marcelo~C Borba}, \bibinfo{person}{Petek
  Askar}, \bibinfo{person}{Johann Engelbrecht}, \bibinfo{person}{George
  Gadanidis}, \bibinfo{person}{Salvador Llinares}, {and}
  \bibinfo{person}{Mario~S{\'a}nchez Aguilar}.}
  \bibinfo{year}{2017}\natexlab{}.
\newblock \showarticletitle{Digital technology in mathematics education:
  Research over the last decade}. In \bibinfo{booktitle}{\emph{Proceedings of
  the 13th International Congress on Mathematical Education}}. Springer, Cham,
  \bibinfo{pages}{221--233}.
\newblock


\bibitem[Bou{\v{s}}ka and Heralov{\'a}(2017)]%
        {bouvska2017opportunities}
\bibfield{author}{\bibinfo{person}{Robert Bou{\v{s}}ka} {and}
  \bibinfo{person}{Ren{\'a}ta~Schneiderov{\'a} Heralov{\'a}}.}
  \bibinfo{year}{2017}\natexlab{}.
\newblock \showarticletitle{Opportunities for use of advanced visualization
  techniques for project coordination}.
\newblock \bibinfo{journal}{\emph{Procedia engineering}}  \bibinfo{volume}{196}
  (\bibinfo{year}{2017}), \bibinfo{pages}{1051--1056}.
\newblock


\bibitem[Bray and Tangney(2017)]%
        {bray2017technology}
\bibfield{author}{\bibinfo{person}{Aibhin Bray} {and} \bibinfo{person}{Brendan
  Tangney}.} \bibinfo{year}{2017}\natexlab{}.
\newblock \showarticletitle{Technology usage in mathematics education
  research--A systematic review of recent trends}.
\newblock \bibinfo{journal}{\emph{Computers \& Education}}
  \bibinfo{volume}{114} (\bibinfo{year}{2017}), \bibinfo{pages}{255--273}.
\newblock


\bibitem[Chaturvedi et~al\mbox{.}(2022)]%
        {chaturvedi2022differential}
\bibfield{author}{\bibinfo{person}{Ramesh~Kumar Chaturvedi},
  \bibinfo{person}{Vishal Verma}, {and} \bibinfo{person}{Kushendra Mishra}.}
  \bibinfo{year}{2022}\natexlab{}.
\newblock \showarticletitle{Differential effect of pre and post cognitive
  skills training program: a study on healthy young children}.
\newblock \bibinfo{journal}{\emph{Smart Learning Environments}}
  \bibinfo{volume}{9}, \bibinfo{number}{1} (\bibinfo{year}{2022}),
  \bibinfo{pages}{1--19}.
\newblock


\bibitem[De~Troyer et~al\mbox{.}(2019)]%
        {de2019truebiters}
\bibfield{author}{\bibinfo{person}{Olga De~Troyer}, \bibinfo{person}{Renny
  Lindberg}, {and} \bibinfo{person}{Pejman Sajjadi}.}
  \bibinfo{year}{2019}\natexlab{}.
\newblock \showarticletitle{TrueBiters, an educational game to practice the
  truth tables of propositional logic: Development, evaluation, and lessons
  learned}.
\newblock \bibinfo{journal}{\emph{Smart Learning Environments}}
  \bibinfo{volume}{6}, \bibinfo{number}{1} (\bibinfo{year}{2019}),
  \bibinfo{pages}{1--17}.
\newblock


\bibitem[Donevska-Todorova(2018)]%
        {donevska2018fostering}
\bibfield{author}{\bibinfo{person}{Ana Donevska-Todorova}.}
  \bibinfo{year}{2018}\natexlab{}.
\newblock \showarticletitle{Fostering students’ competencies in linear
  algebra with digital resources}.
\newblock In \bibinfo{booktitle}{\emph{Challenges and Strategies in Teaching
  Linear Algebra}}. \bibinfo{publisher}{Springer}, \bibinfo{pages}{261--276}.
\newblock


\bibitem[Fadiran et~al\mbox{.}(2018)]%
        {fadiran2018can}
\bibfield{author}{\bibinfo{person}{Olakumbi~A Fadiran}, \bibinfo{person}{Judy
  Van~Biljon}, {and} \bibinfo{person}{Marthie~A Schoeman}.}
  \bibinfo{year}{2018}\natexlab{}.
\newblock \showarticletitle{How can visualisation principles be used to support
  knowledge transfer in teaching and learning?}. In
  \bibinfo{booktitle}{\emph{2018 Conference on Information Communications
  Technology and Society (ICTAS)}}. IEEE, \bibinfo{pages}{1--6}.
\newblock


\bibitem[Fernandes and Borkar(2013)]%
        {fernandes2013application}
\bibfield{author}{\bibinfo{person}{Chilton Fernandes} {and}
  \bibinfo{person}{Samarth Borkar}.} \bibinfo{year}{2013}\natexlab{}.
\newblock \showarticletitle{Application of Vedic Mathematics in Computer
  Architecture}.
\newblock \bibinfo{journal}{\emph{International Journal of Research in
  Engineering and Science (IJRES)}} \bibinfo{volume}{1}, \bibinfo{number}{5}
  (\bibinfo{year}{2013}), \bibinfo{pages}{40--45}.
\newblock


\bibitem[F{\i}rat and Laramee(2018)]%
        {firat2018towards}
\bibfield{author}{\bibinfo{person}{Elif~E F{\i}rat} {and}
  \bibinfo{person}{Robert~S Laramee}.} \bibinfo{year}{2018}\natexlab{}.
\newblock \showarticletitle{Towards a survey of interactive visualization for
  education}.
\newblock \bibinfo{journal}{\emph{EG UK Computer Graphics \& Visual Computing,
  Eurographics Proceedings}} (\bibinfo{year}{2018}).
\newblock


\bibitem[Fuchsova and Korenova(2019)]%
        {fuchsova2019visualisation}
\bibfield{author}{\bibinfo{person}{Maria Fuchsova} {and} \bibinfo{person}{Lilla
  Korenova}.} \bibinfo{year}{2019}\natexlab{}.
\newblock \showarticletitle{Visualisation in Basic Science and Engineering
  Education of Future Primary School Teachers in Human Biology Education Using
  Augmented Reality.}
\newblock \bibinfo{journal}{\emph{European Journal of Contemporary Education}}
  \bibinfo{volume}{8}, \bibinfo{number}{1} (\bibinfo{year}{2019}),
  \bibinfo{pages}{92--102}.
\newblock


\bibitem[Geng and Yamada(2020)]%
        {geng2020augmented}
\bibfield{author}{\bibinfo{person}{Xuewang Geng} {and}
  \bibinfo{person}{Masanori Yamada}.} \bibinfo{year}{2020}\natexlab{}.
\newblock \showarticletitle{An augmented reality learning system for Japanese
  compound verbs: study of learning performance and cognitive load}.
\newblock \bibinfo{journal}{\emph{Smart Learning Environments}}
  \bibinfo{volume}{7}, \bibinfo{number}{1} (\bibinfo{year}{2020}),
  \bibinfo{pages}{1--19}.
\newblock


\bibitem[Gilbert(2005)]%
        {gilbert2005visualization}
\bibfield{author}{\bibinfo{person}{John~K Gilbert}.}
  \bibinfo{year}{2005}\natexlab{}.
\newblock \showarticletitle{Visualization: A metacognitive skill in science and
  science education}.
\newblock In \bibinfo{booktitle}{\emph{Visualization in science education}}.
  \bibinfo{publisher}{Springer}, \bibinfo{pages}{9--27}.
\newblock


\bibitem[G{\"u}ven and Kosa(2008)]%
        {guven2008effect}
\bibfield{author}{\bibinfo{person}{B{\"u}lent G{\"u}ven} {and}
  \bibinfo{person}{Temel Kosa}.} \bibinfo{year}{2008}\natexlab{}.
\newblock \showarticletitle{The effect of dynamic geometry software on student
  mathematics teachers' spatial visualization skills.}
\newblock \bibinfo{journal}{\emph{Turkish Online Journal of Educational
  Technology-TOJET}} \bibinfo{volume}{7}, \bibinfo{number}{4}
  (\bibinfo{year}{2008}), \bibinfo{pages}{100--107}.
\newblock


\bibitem[Hadjerrouit and Gautestad(2018)]%
        {hadjerrouit2018using}
\bibfield{author}{\bibinfo{person}{Said Hadjerrouit} {and}
  \bibinfo{person}{Harald~H Gautestad}.} \bibinfo{year}{2018}\natexlab{}.
\newblock \showarticletitle{Using the visualization tool SimReal to orchestrate
  mathematical teaching for engineering students}. In
  \bibinfo{booktitle}{\emph{2018 IEEE Global Engineering Education Conference
  (EDUCON)}}. IEEE, \bibinfo{pages}{38--42}.
\newblock


\bibitem[Huddar et~al\mbox{.}(2013)]%
        {huddar2013novel}
\bibfield{author}{\bibinfo{person}{Sushma~R Huddar},
  \bibinfo{person}{Sudhir~Rao Rupanagudi}, \bibinfo{person}{M Kalpana}, {and}
  \bibinfo{person}{Surabhi Mohan}.} \bibinfo{year}{2013}\natexlab{}.
\newblock \showarticletitle{Novel high speed vedic mathematics multiplier using
  compressors}. In \bibinfo{booktitle}{\emph{2013 International
  Mutli-Conference on Automation, Computing, Communication, Control and
  Compressed Sensing (iMac4s)}}. IEEE, \bibinfo{pages}{465--469}.
\newblock


\bibitem[Ineson and Babbar(2020)]%
        {ineson2020mental}
\bibfield{author}{\bibinfo{person}{Gwen Ineson} {and} \bibinfo{person}{Sunita
  Babbar}.} \bibinfo{year}{2020}\natexlab{}.
\newblock \showarticletitle{Mental Maths: Just about what we do in our heads?}
\newblock In \bibinfo{booktitle}{\emph{Debates in Mathematics Education}}.
  \bibinfo{publisher}{Routledge}, \bibinfo{pages}{169--181}.
\newblock


\bibitem[Jacquesson et~al\mbox{.}(2020)]%
        {jacquesson2020stereoscopic}
\bibfield{author}{\bibinfo{person}{Timoth{\'e}e Jacquesson},
  \bibinfo{person}{Emile Simon}, \bibinfo{person}{Corentin Dauleac},
  \bibinfo{person}{Lo{\"\i}c Margueron}, \bibinfo{person}{Philip Robinson},
  {and} \bibinfo{person}{Patrick Mertens}.} \bibinfo{year}{2020}\natexlab{}.
\newblock \showarticletitle{Stereoscopic three-dimensional visualization:
  interest for neuroanatomy teaching in medical school}.
\newblock \bibinfo{journal}{\emph{Surgical and Radiologic Anatomy}}
  (\bibinfo{year}{2020}), \bibinfo{pages}{1--9}.
\newblock


\bibitem[Kamalodeen et~al\mbox{.}(2021)]%
        {kamalodeen2021designing}
\bibfield{author}{\bibinfo{person}{Vimala~Judy Kamalodeen},
  \bibinfo{person}{Nalini Ramsawak-Jodha}, \bibinfo{person}{Sandra
  Figaro-Henry}, \bibinfo{person}{Sharon~J Jaggernauth}, {and}
  \bibinfo{person}{Zhanna Dedovets}.} \bibinfo{year}{2021}\natexlab{}.
\newblock \showarticletitle{Designing gamification for geometry in elementary
  schools: insights from the designers}.
\newblock \bibinfo{journal}{\emph{Smart Learning Environments}}
  \bibinfo{volume}{8}, \bibinfo{number}{1} (\bibinfo{year}{2021}),
  \bibinfo{pages}{1--24}.
\newblock


\bibitem[Kang et~al\mbox{.}(2020)]%
        {kang2020impact}
\bibfield{author}{\bibinfo{person}{Keegan Kang}, \bibinfo{person}{Sergey
  Kushnarev}, \bibinfo{person}{Wong~Wei Pin}, \bibinfo{person}{Omar Ortiz},
  {and} \bibinfo{person}{Jacob~Chen Shihang}.} \bibinfo{year}{2020}\natexlab{}.
\newblock \showarticletitle{Impact of Virtual Reality on the Visualization of
  Partial Derivatives in a Multivariable Calculus Class}.
\newblock \bibinfo{journal}{\emph{IEEE Access}}  \bibinfo{volume}{8}
  (\bibinfo{year}{2020}), \bibinfo{pages}{58940--58947}.
\newblock


\bibitem[Kaufmann(2002)]%
        {kaufmann2002construct3d}
\bibfield{author}{\bibinfo{person}{Hannes Kaufmann}.}
  \bibinfo{year}{2002}\natexlab{}.
\newblock \showarticletitle{Construct3D: an augmented reality application for
  mathematics and geometry education}. In \bibinfo{booktitle}{\emph{Proceedings
  of the tenth ACM international conference on Multimedia}}.
  \bibinfo{pages}{656--657}.
\newblock


\bibitem[Kodali et~al\mbox{.}(2015)]%
        {kodali2015fpga}
\bibfield{author}{\bibinfo{person}{Ravi~Kishore Kodali},
  \bibinfo{person}{Lakshmi Boppana}, {and} \bibinfo{person}{Sai~Sourabh
  Yenamachintala}.} \bibinfo{year}{2015}\natexlab{}.
\newblock \showarticletitle{FPGA implementation of vedic floating point
  multiplier}. In \bibinfo{booktitle}{\emph{2015 IEEE International Conference
  on Signal Processing, Informatics, Communication and Energy Systems
  (SPICES)}}. IEEE, \bibinfo{pages}{1--4}.
\newblock


\bibitem[K{\"o}lling(2010)]%
        {kolling2010greenfoot}
\bibfield{author}{\bibinfo{person}{Michael K{\"o}lling}.}
  \bibinfo{year}{2010}\natexlab{}.
\newblock \showarticletitle{The greenfoot programming environment}.
\newblock \bibinfo{journal}{\emph{ACM Transactions on Computing Education
  (TOCE)}} \bibinfo{volume}{10}, \bibinfo{number}{4} (\bibinfo{year}{2010}),
  \bibinfo{pages}{1--21}.
\newblock


\bibitem[Liapi(2003)]%
        {liapi20034d}
\bibfield{author}{\bibinfo{person}{Katherine~A Liapi}.}
  \bibinfo{year}{2003}\natexlab{}.
\newblock \showarticletitle{4D visualization of highway construction projects}.
  In \bibinfo{booktitle}{\emph{Proceedings on Seventh International Conference
  on Information Visualization, 2003. IV 2003.}} IEEE,
  \bibinfo{pages}{639--644}.
\newblock


\bibitem[Lo and Hew(2020)]%
        {lo2020comparison}
\bibfield{author}{\bibinfo{person}{Chung~Kwan Lo} {and}
  \bibinfo{person}{Khe~Foon Hew}.} \bibinfo{year}{2020}\natexlab{}.
\newblock \showarticletitle{A comparison of flipped learning with gamification,
  traditional learning, and online independent study: the effects on
  students’ mathematics achievement and cognitive engagement}.
\newblock \bibinfo{journal}{\emph{Interactive Learning Environments}}
  \bibinfo{volume}{28}, \bibinfo{number}{4} (\bibinfo{year}{2020}),
  \bibinfo{pages}{464--481}.
\newblock


\bibitem[Misrom et~al\mbox{.}(2020)]%
        {misrom2020enhancing}
\bibfield{author}{\bibinfo{person}{Noor~Binti Misrom},
  \bibinfo{person}{Abdurrahman Muhammad}, \bibinfo{person}{Abdul Abdullah},
  \bibinfo{person}{Sharifah Osman}, \bibinfo{person}{Mohd Hamzah}, {and}
  \bibinfo{person}{Ahmad Fauzan}.} \bibinfo{year}{2020}\natexlab{}.
\newblock \showarticletitle{Enhancing students’ higher-order thinking skills
  (HOTS) through an inductive reasoning strategy using geogebra}.
\newblock \bibinfo{journal}{\emph{International Journal of Emerging
  Technologies in Learning (iJET)}} \bibinfo{volume}{15}, \bibinfo{number}{3}
  (\bibinfo{year}{2020}), \bibinfo{pages}{156--179}.
\newblock


\bibitem[Motschnig et~al\mbox{.}(2016)]%
        {motschnig2016team}
\bibfield{author}{\bibinfo{person}{Renate Motschnig}, \bibinfo{person}{Michael
  Sedlmair}, \bibinfo{person}{Svenja Schr{\"o}der}, {and}
  \bibinfo{person}{Torsten M{\"o}ller}.} \bibinfo{year}{2016}\natexlab{}.
\newblock \showarticletitle{A team-approach to putting learner-centered
  principles to practice in a large course on Human-Computer Interaction}. In
  \bibinfo{booktitle}{\emph{2016 IEEE Frontiers in Education Conference
  (FIE)}}. IEEE, \bibinfo{pages}{1--9}.
\newblock


\bibitem[Mudaly and Rampersad(2010)]%
        {mudaly2010role}
\bibfield{author}{\bibinfo{person}{Vimolan Mudaly} {and}
  \bibinfo{person}{Rajesh Rampersad}.} \bibinfo{year}{2010}\natexlab{}.
\newblock \showarticletitle{The role of visualisation in learners' conceptual
  understanding of graphical functional relationships}.
\newblock \bibinfo{journal}{\emph{African Journal of Research in Mathematics,
  Science and Technology Education}} \bibinfo{volume}{14}, \bibinfo{number}{1}
  (\bibinfo{year}{2010}), \bibinfo{pages}{36--48}.
\newblock


\bibitem[Nicolson(1990)]%
        {nicolson1990design}
\bibfield{author}{\bibinfo{person}{Roderick~I Nicolson}.}
  \bibinfo{year}{1990}\natexlab{}.
\newblock \showarticletitle{Design and evaluation of the SUMIT intelligent
  teaching assistant for arithmetic}.
\newblock \bibinfo{journal}{\emph{Interactive Learning Environments}}
  \bibinfo{volume}{1}, \bibinfo{number}{4} (\bibinfo{year}{1990}),
  \bibinfo{pages}{265--287}.
\newblock


\bibitem[Parish and Edmondson(2019)]%
        {parish2019data}
\bibfield{author}{\bibinfo{person}{Chad~M Parish} {and}
  \bibinfo{person}{Philip~D Edmondson}.} \bibinfo{year}{2019}\natexlab{}.
\newblock \showarticletitle{Data visualization heuristics for the physical
  sciences}.
\newblock \bibinfo{journal}{\emph{Materials \& Design}}  \bibinfo{volume}{179}
  (\bibinfo{year}{2019}), \bibinfo{pages}{107868}.
\newblock


\bibitem[Presmeg(2020)]%
        {presmeg2020visualization}
\bibfield{author}{\bibinfo{person}{Norma Presmeg}.}
  \bibinfo{year}{2020}\natexlab{}.
\newblock \showarticletitle{Visualization and learning in mathematics
  education}.
\newblock \bibinfo{journal}{\emph{Encyclopedia of mathematics education}}
  (\bibinfo{year}{2020}), \bibinfo{pages}{900--904}.
\newblock


\bibitem[Raikhola and Campus(2020)]%
        {raikhola2020thematic}
\bibfield{author}{\bibinfo{person}{Sher~Singh Raikhola} {and}
  \bibinfo{person}{Valmeeki Campus}.} \bibinfo{year}{2020}\natexlab{}.
\newblock \showarticletitle{A Thematic Analysis on Vedic Mathematics and Its
  Importance}.
\newblock \bibinfo{journal}{\emph{Open Access Library Journal}}
  \bibinfo{volume}{7}, \bibinfo{number}{08} (\bibinfo{year}{2020}),
  \bibinfo{pages}{1}.
\newblock


\bibitem[Shatri and Buza(2017)]%
        {shatri2017use}
\bibfield{author}{\bibinfo{person}{Kyvete Shatri} {and}
  \bibinfo{person}{Kastriot Buza}.} \bibinfo{year}{2017}\natexlab{}.
\newblock \showarticletitle{The use of visualization in teaching and learning
  process for developing critical thinking of students}.
\newblock \bibinfo{journal}{\emph{European Journal of Social Science Education
  and Research}} \bibinfo{volume}{4}, \bibinfo{number}{1}
  (\bibinfo{year}{2017}), \bibinfo{pages}{71--74}.
\newblock


\bibitem[Shaziya and Zaheer(2021)]%
        {shaziya2021strategies}
\bibfield{author}{\bibinfo{person}{Humera Shaziya} {and}
  \bibinfo{person}{Raniah Zaheer}.} \bibinfo{year}{2021}\natexlab{}.
\newblock \showarticletitle{Strategies to Effectively Integrate Visualization
  with Active Learning in Computer Science Class}. In
  \bibinfo{booktitle}{\emph{Proceedings of International Conference on
  Computational Intelligence and Data Engineering}}. Springer,
  \bibinfo{pages}{69--81}.
\newblock


\bibitem[Shriki and Lavy(2018)]%
        {shriki2018engagement}
\bibfield{author}{\bibinfo{person}{Atara Shriki} {and} \bibinfo{person}{Ilana
  Lavy}.} \bibinfo{year}{2018}\natexlab{}.
\newblock \showarticletitle{Engagement in Vedic Mathematics as means for
  strengthening self-efficacy of low achievers}. In
  \bibinfo{booktitle}{\emph{Proceedings of EDULEARN18 Conference, 10th annual
  International Conference on Education and New Learning Technologies}}.
  \bibinfo{pages}{5441--5449}.
\newblock


\bibitem[Teichrew and Erb(2020)]%
        {teichrew2020augmented}
\bibfield{author}{\bibinfo{person}{Albert Teichrew} {and}
  \bibinfo{person}{Roger Erb}.} \bibinfo{year}{2020}\natexlab{}.
\newblock \showarticletitle{How augmented reality enhances typical classroom
  experiments: examples from mechanics, electricity and optics}.
\newblock \bibinfo{journal}{\emph{Physics Education}} \bibinfo{volume}{55},
  \bibinfo{number}{6} (\bibinfo{year}{2020}), \bibinfo{pages}{065029}.
\newblock


\bibitem[Tirtha and Agrawala(1992)]%
        {tirtha1992vedic}
\bibfield{author}{\bibinfo{person}{Swami Bharati~Krishna Tirtha} {and}
  \bibinfo{person}{Agrawala}.} \bibinfo{year}{1992}\natexlab{}.
\newblock \bibinfo{booktitle}{\emph{Vedic mathematics}}.
  Vol.~\bibinfo{volume}{10}.
\newblock \bibinfo{publisher}{Motilal Banarsidass Publ.}
\newblock


\bibitem[Trachtenberg(2011)]%
        {trachtenberg2011trachtenberg}
\bibfield{author}{\bibinfo{person}{Jakow Trachtenberg}.}
  \bibinfo{year}{2011}\natexlab{}.
\newblock \bibinfo{booktitle}{\emph{The Trachtenberg speed system of basic
  mathematics}}.
\newblock \bibinfo{publisher}{Souvenir Press}.
\newblock


\bibitem[Venkatesh and Bala(2008)]%
        {venkatesh2008technology}
\bibfield{author}{\bibinfo{person}{Viswanath Venkatesh} {and}
  \bibinfo{person}{Hillol Bala}.} \bibinfo{year}{2008}\natexlab{}.
\newblock \showarticletitle{Technology acceptance model 3 and a research agenda
  on interventions}.
\newblock \bibinfo{journal}{\emph{Decision sciences}} \bibinfo{volume}{39},
  \bibinfo{number}{2} (\bibinfo{year}{2008}), \bibinfo{pages}{273--315}.
\newblock


\bibitem[Yang et~al\mbox{.}(2018)]%
        {yang2018evaluations}
\bibfield{author}{\bibinfo{person}{Jeong Yang}, \bibinfo{person}{Young Lee},
  {and} \bibinfo{person}{Kai~H Chang}.} \bibinfo{year}{2018}\natexlab{}.
\newblock \showarticletitle{Evaluations of JaguarCode: A web-based
  object-oriented programming environment with static and dynamic
  visualization}.
\newblock \bibinfo{journal}{\emph{Journal of Systems and Software}}
  \bibinfo{volume}{145} (\bibinfo{year}{2018}), \bibinfo{pages}{147--163}.
\newblock


\bibitem[Ye et~al\mbox{.}(2007)]%
        {ye2007enhancing}
\bibfield{author}{\bibinfo{person}{En Ye}, \bibinfo{person}{Chang Liu}, {and}
  \bibinfo{person}{Jennifer~A Polack-Wahl}.} \bibinfo{year}{2007}\natexlab{}.
\newblock \showarticletitle{Enhancing software engineering education using
  teaching aids in 3-D online virtual worlds}. In
  \bibinfo{booktitle}{\emph{2007 37th Annual Frontiers In Education
  Conference-Global Engineering: Knowledge Without Borders, Opportunities
  Without Passports}}. IEEE, \bibinfo{pages}{T1E--8}.
\newblock


\end{thebibliography}










\end{document}